\documentclass[twoside]{LCWS11}
\usepackage[latin1]{inputenc}
\usepackage[dvips]{graphicx,epsfig,color}
\usepackage{wrapfig,rotating}
\usepackage{amssymb,amsmath,array}
\usepackage{cite}
\usepackage{url}
\begin{document}
\def\ttbar{\ensuremath{t\bar{t}}\,}
\def\xsectt{\ensuremath{\sigma_{ t\bar{t}}}\,}
\def\xsect{\ensuremath{\sigma_{t}}\,}
\def\MET{\ensuremath{E_{\mathrm{T}}^{\mathrm{miss}}}\,}
\def\pt{\ensuremath{p_{\mathrm{T}}}\,}
\def\ilumi{\ensuremath{\int Ldt}\,}
\def\ifb{\ensuremath{\mathrm{fb^{-1}}}\,}
\def\ipb{\ensuremath{\mathrm{pb^{-1}}}\,}
\def\mtop{\ensuremath{m_{t}}\,}
\title{
Top-Quark Physics Results From LHC} 
\author{Luca Fiorini$^1$\\
on behalf of the ATLAS and CMS collaborations.
\vspace{.3cm}\\
1- Instituto de F\'isica Corpuscular (IFIC), Universitat de Val\`encia and CSIC.\\
Edificio Institutos de Investigaci\'on
Apartado de Correos 22085 \\
46071 Valencia - Spain \\
E-mail: \texttt{Luca.Fiorini@cern.ch}
}

\maketitle

\begin{abstract}
The top-quark is a fundamental element of the physics program at the Large Hadron Collider (LHC).  
We review the current status of the top-quark measurements performed by ATLAS and CMS experiments in pp collisions at $\sqrt{s}$=7 TeV by presenting the recent results of the top-quark production rates, top mass measurements and additional top quark properties.\\
We will also describe the recent searches for physics beyond the Standard Model in the top-quark sector.
\end{abstract}

\section{Introduction}
The top-quark is the heaviest known elementary particle, with a mass measured by TeVatron experiments to be about 173 GeV~\cite{toppdg}. 
Due to its high mass, the top-quark is believed to play a special role
in the electro-weak symmetry breaking mechanism and possibly in models of new physics 
beyond the Standard Model (SM).

The Large Hadron Collider (LHC)~\cite{lhc} operated in 2010 and 2011 with an energy in the centre of mass ($\sqrt{s}$) of 7 TeV. The delivered integrated luminosity ($\int L dt$) to ATLAS~\cite{atlas} and CMS~\cite{cms} experiments was of about 5~\ifb per experiment in 2011 and the peak luminosity was of $3.7\cdot 10^{33}$ cm$^{-2}$s$^{-1}$. Measurements presented in this proceeding are based on the statistics collected up to summer 2011 and use at most 2.1~\ifb.
The \ttbar\ production rate at LHC $\sqrt{s}$~=~7 TeV is a factor of 20
larger than at the TeVatron, allowing the production of top-quarks with unprecedented abundance.

\section{Top-quark production measurements}
\subsection{\textbf{\ttbar} production measurements}
Figure \ref{fig:tt-prod} shows the leading order diagrams of the \ttbar\ 
production process. The \ttbar\ production is dominated by the gluon-fusion process at LHC energies. 
The \ttbar\ production cross-section, \xsectt, is predicted to be
 at the approximate next to next to leading order $\xsectt^{approx~ NNLO}$ =
$165^{+11}_{-16}$ pb~\cite{xsec}. 
The top-quark decays to a $W$ and a $b$-quark almost 100\% of the times. The $W$ decays hadronically in about 68\% of the times. The \ttbar\ final states are categorised by the number of leptons from the $W$ decays in the final state: di-lepton, single lepton and full hadronic channels.
Both ATLAS and CMS measured \xsectt\  in several final states: di-lepton ($\ell=e,\mu,\tau$), single-lepton and full hadronic. The results are shown in Figures~\ref{fig:xsec-atlas} and \ref{fig:xsec-cms}.
\begin{itemize}
\item The \ttbar\ cross-section has been measured by ATLAS~\cite{atlassinglelep,atlassinglelepnew2} and CMS~\cite{cmssinglelep} in the  single-lepton channel with either an electron or a muon in the final state. The events are required to have a high-\pt\ lepton and at least three jets. The \xsectt\  has been measured with and without the requirement of a $b$-tagged jet. Results are obtained for \ilumi= 35 \ipb and 0.7~\ifb for ATLAS and (0.8-1.1)~\ifb for CMS. Figure~\ref{fig:xsec-atlas} show the ATLAS single-lepton results obtained with  \ilumi= 35 \ipb, while result for 0.7~\ifb is  \xsectt~=~179~$\pm$~9.8~(stat.)~$\pm$~9.7~(syst.)~$\pm$~6.6~(lumi.)~pb. Figure~\ref{fig:xsec-cms} shows the CMS results with 2011 data. Uncertainties with 2011 data are at the level of 9\% for ATLAS and CMS which is comparable to the NNLO theoretical uncertainty.\\
\item The event selection for di-lepton final states require the presence of 2 high-\pt leptons of opposite charge, two central jets (from $b$-quarks) and large \MET or large transverse activity ($H_{T}$). The \MET is the missing energy in the transverse plane, calculated taking into account the transverse momentum of the muons and transverse energy of the electrons and jets in the events. $H_{T}$ is the scalar sum of the transverse momentum of the muons, transverse energy of the electrons, \MET and transverse energy of the jets in the event. 
In the case that the leptons in the final state are $e$ or $\mu$, ATLAS results are obtained with and without the requirement of a $b$-tagged jet~\cite{atlasdileptag,atlasdilepnew}; CMS results are obtained with $b$-tagging requirement~\cite{cmsdilep}. ATLAS produced results with 0.7~\ifb of data, while CMS used 1.1~\ifb of data for the result with $b$-tagging requirement. Their uncertainties are about 11\% dominated by systematics.
The \xsectt\  has been measured also in the $\tau\mu$ final state by ATLAS~\cite{atlastaumu} and CMS~\cite{cmstaumu}, where a calorimeter-seeded $\tau$ is reconstructed in the event together with a high-\pt muon. The resulting \xsectt\  measurements are obtained for \ilumi=1.1~\ifb and have similar uncertainties: 24\% CMS and 21\% ATLAS\footnote{ATLAS: \xsectt~=~142~$\pm$~21~(stat.)~$^{+20}_{-16}$~(syst.)~$\pm$~5~(lumi.)~pb ; CMS results are in Figure~\ref{fig:xsec-cms}.}. 
\item The \xsectt\  has been measured by ATLAS~\cite{atlasfullhad} and CMS~\cite{cmsfullhad} also in the full hadronic final state. Events are required to have at least six high-\pt\ jets of which at least two $b$-tagged jets. The results are obtained for \ilumi~=~1.1 (1.0)~\ifb for CMS (ATLAS)\footnote{  CMS results are in Figure~\ref{fig:xsec-cms} and 
ATLAS obtains  \xsectt~=~167~$\pm$~18~(stat.)~$\pm$~78~(syst.)~$\pm$~6~(lumi.)~pb.}. 
\end{itemize}

CMS combined the cross-section measurements obtaining  \xsectt~=~166~$\pm$~2~(stat.)~$\pm$ 11~(syst.)~$\pm$~8~(lumi.)~pb.
The measurements of the \ttbar\ cross-section at LHC are in agreement with the theoretical predictions. Their accuracy is similar to the uncertainty of the NNLO prediction and the most sensitive results in the single-lepton channel are limited by systematic uncertainties.

\subsection{Single-top production measurements}
 Figure \ref{fig:stop-prod} shows the leading order diagrams of the electro-weak single-top-quark production, that is characterised by the $W$-mediated t-channel ($\sigma_{t}^{approx~ NNLO}$ =$64 \pm 3$ pb)~\cite{tchanxsec}, s-channel ($\sigma_{s}^{approx~ NNLO}$ =$4.6 \pm 0.3$ pb)~\cite{schanxsec} and the associated $Wt$ production ($\sigma_{Wt}^{approx~ NNLO}$ =$15.7^{+1.3}_{-1.4}$ pb)~\cite{Wtchanxsec}. Compared to TeVatron experiments, the single-top production in t-channel and $Wt$ mode is much larger than the s-channel.

The single-top-quark production in the $t$-channel has been measured by ATLAS~\cite{atlastchan} with 0.7~\ifb of data and by CMS~\cite{cmstchan} with 36~\ipb of data. ATLAS measurement is \xsect~=~90~$\pm$9~(stat.)~$^{+31}_{-20}$~(syst.)~pb and CMS measurement is \xsect~=~84~$\pm$30~(stat.+syst.) $\pm$3~(lumi.)~pb.
Searches of the $Wt$ channel have been performed by CMS~\cite{cmsWt} and ATLAS~\cite{atlasWt} with 2.1~\ifb and 0.7~\ifb of data respectively. This mode has escaped so far direct observation. ATLAS rejects the background-only hypothesis at the 1.2 $\sigma$ level and CMS obtains a signal significance of 2.7~$\sigma$. 
Searches of single-top-quark production in the $s$-channel have been performed by ATLAS with  \ilumi~=~0.7~\ifb~\cite{atlasschan}. An  upper limit at 95\% C.L. on the production cross section of $\sigma_{s}< 26.5$ pb has been obtained.

\begin{figure}
\begin{minipage}[b]{0.49\linewidth}
\centerline{\includegraphics[width=0.95\columnwidth]{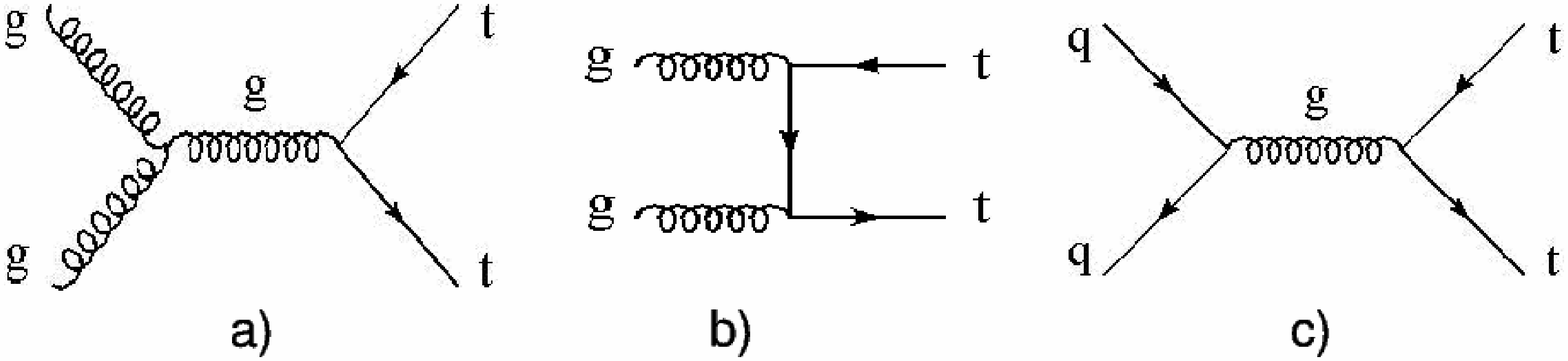}}
\caption{ Lowest level diagrams of the \ttbar production. Gluon scattering processes, {\tt a)} and {\tt b)}, are the
    dominant processes at LHC energies, while quark scattering, process {\tt c)}, is the dominant
    one at TeVatron energies.}\label{fig:tt-prod}
\end{minipage}
\hspace{0.5cm}
\begin{minipage}[b]{0.49\linewidth}
\centerline{\includegraphics[width=0.95\columnwidth]{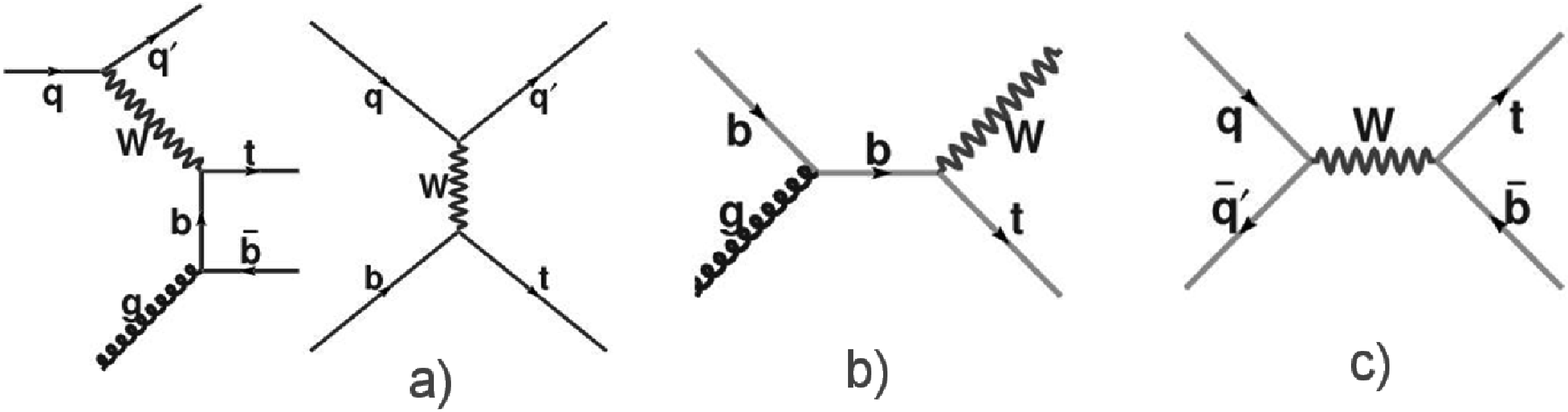}}
\caption{  Diagrams of single-top production at the lowest
    level: {\tt a)} t-channel, {\tt b)} Wt associated production, 
{\tt c)} s-channel.}\label{fig:stop-prod}
\end{minipage}
\end{figure}


\begin{figure}
\begin{minipage}[b]{0.49\linewidth}
\centerline{\includegraphics[width=0.95\columnwidth]{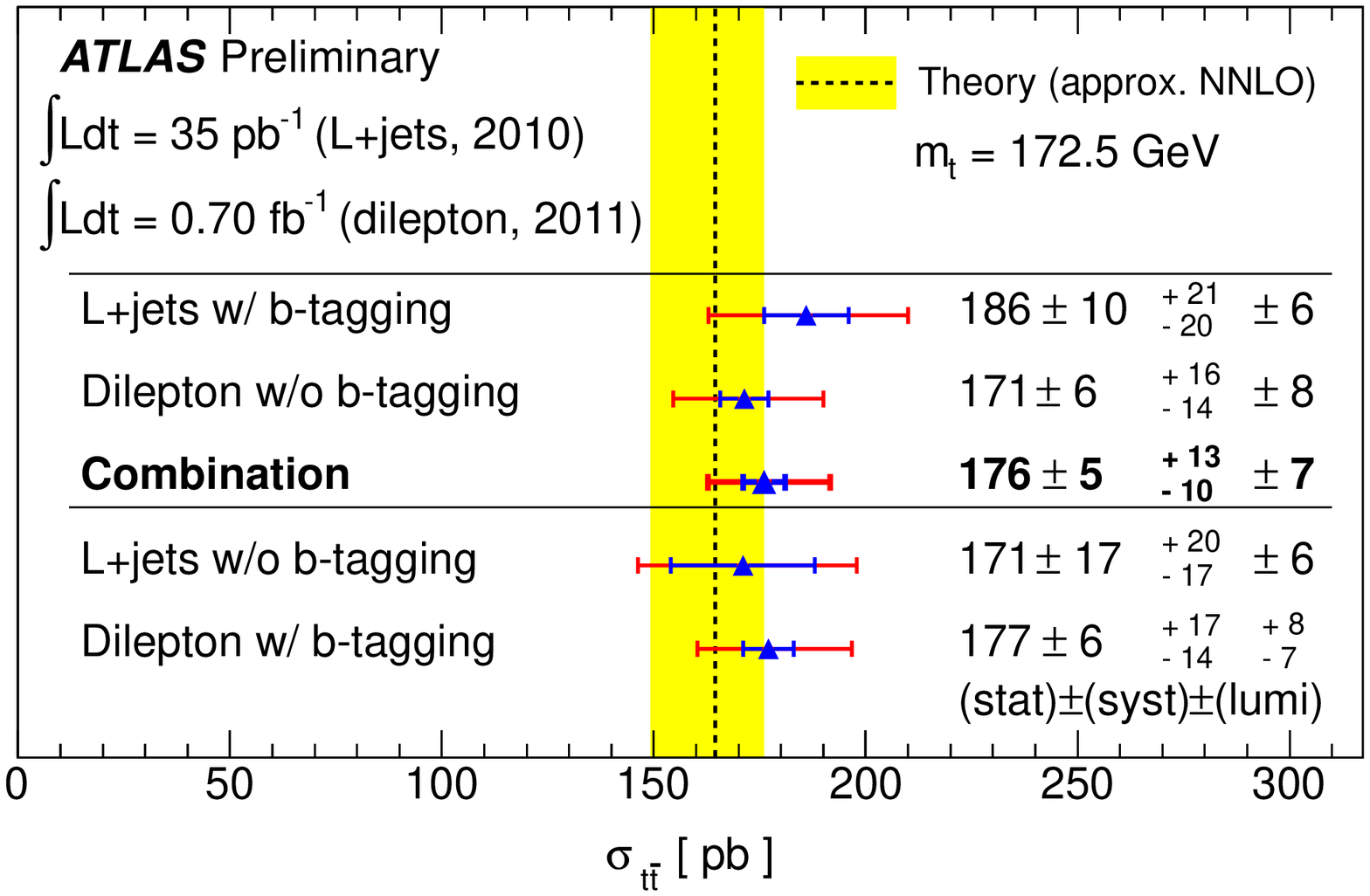}}
\caption{ Summary of the \ttbar\ cross-section measurements from ATLAS Collaboration~\cite{atlassinglelepnew}.}\label{fig:xsec-atlas}
\end{minipage}
\hspace{0.5cm}
\begin{minipage}[b]{0.49\linewidth}
\centerline{\includegraphics[width=0.95\columnwidth]{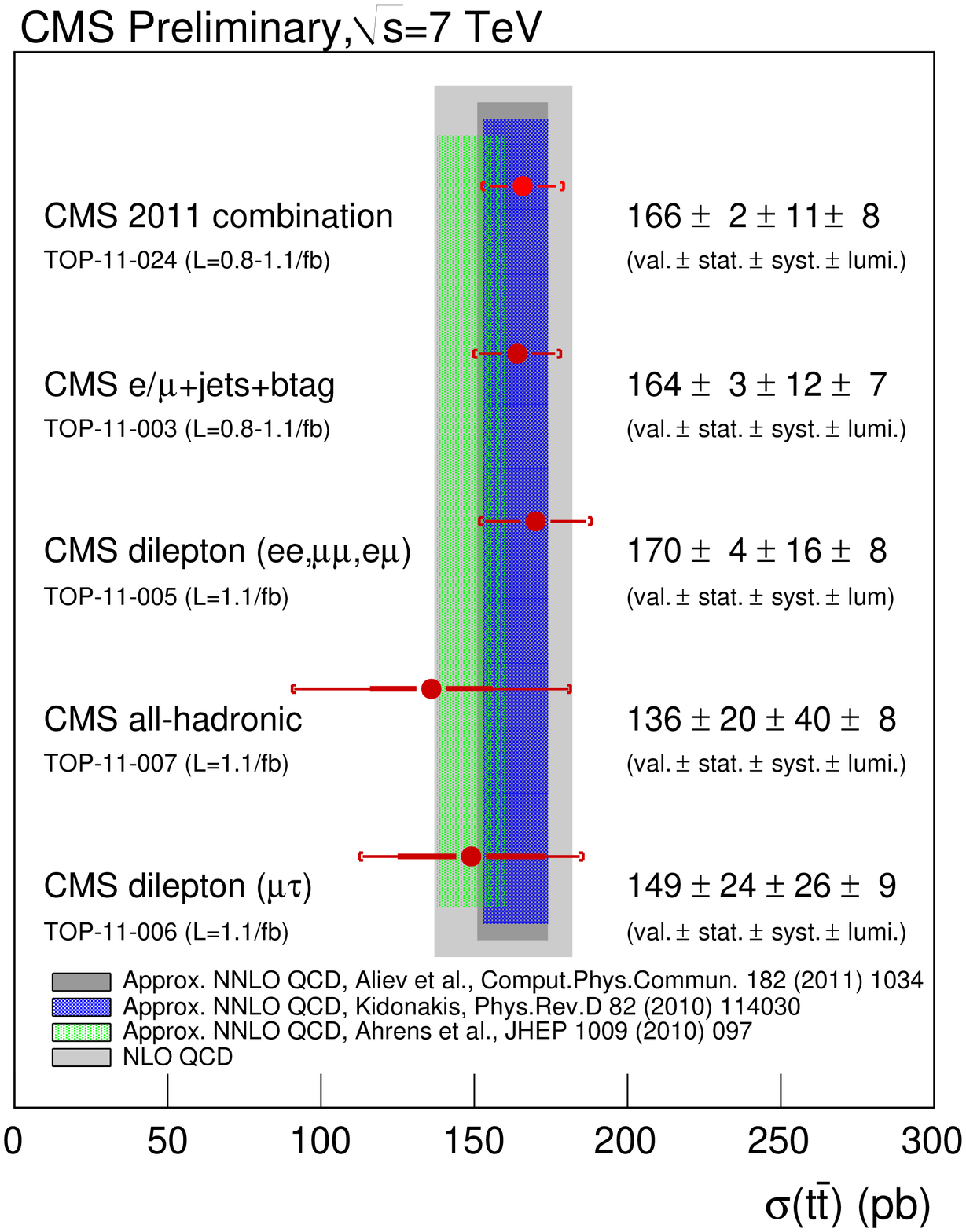}}
\caption{ Summary of the \ttbar\ cross-section measurements from CMS Collaboration~\cite{cmssinglelepnew}.}\label{fig:xsec-cms}
\end{minipage}
\end{figure}

\section{Top-quark properties}
\subsection{Top-quark mass}
Among the various top-quark properties, ATLAS and CMS measured the top-quark mass (\mtop) in several final states.\\
In the SM the mass of the top-quark is a free parameter that, together with the $W$ mass, can constrain the mass value of an eventual Higgs boson~\cite{ewfit}.
Figure~\ref{fig:mass-atlas} summarises the measurement of \mtop\ by ATLAS experiment~\cite{atlasmass}. A measurement in the single-lepton final state has been obtained with~0.7~\ifb of data with an uncertainty that is at the level of 1.6\%:  \mtop~=~175.9~$\pm$~0.9~(stat.)~$\pm$~2.7~(syst.)~GeV. CMS measured \mtop\ with  \ilumi~=~36~\ipb in the single-lepton~\cite{cmsmasssinglelep} and di-lepton~\cite{cmsmassdilep} final states, obtaining \mtop~=~173.1~$\pm$ 2.1~(stat.)~$^{+2.8}_{-2.1}$~(syst.)~GeV (single-lepton) and \mtop~=~175.5~$\pm$~4.6~(stat.)~$\pm$~4.6~(syst.)~GeV (di-lepton).
Both experiments have  also measured the top-quark pole mass, $m_{pole}$, from the \xsectt\ measurement~\cite{atlasmpole}, \cite{cmsmpole}. This is a complementary measurement of a different observable with respect to the measurement of the top-quark invariant mass from the top-quark decay products. Results are shown in Figure~\ref{fig:mass-cms}.

\subsection{Top-anti-top quark mass difference}
The SM assumes that top and anti-top-quarks have the same mass. An eventual CPT violation can manifest itself as a mass difference between the top and anti-top-quark. CMS measured the mass difference in the $\mu$+jets final state~\cite{cmsmassdiff}: $\Delta$\mtop~=~-1.2~$\pm$~1.2~(stat.)~$\pm$ 0.5~(syst.)~GeV. These measurements are threfore consistent with the SM.

\subsection{Top-quark charge asymmetry}
Recently TeVatron experiments reported forward-backward asymmetry in \ttbar\ events that is larger then theoretical predictions by about 3$\sigma$~\cite{d0afb,cdfafb}. At the Next-to-Leading-Order (NLO), a small difference of about 1\% in the top and anti-top rapidity distribution is expected at LHC~\cite{theoafb}. Results are obtained for CMS~\cite{cmsafb} and ATLAS~\cite{atlasafb} in the single-lepton final state for  \ilumi~=~1.1 (0.7)~\ifb for CMS (ATLAS) experiment. CMS measured the asymmetry from the top-anti-top pseudo-rapidity distribution, while ATLAS measured it from the rapidity distribution. The asymmetry measured is $A_{C}$~=~-1.6~$\pm$~3.0~$^{+1.0}_{-1.9} \%$ (CMS) for a theoretical prediction of 1.3\% and  $A_{C}$~=~-2.4~$\pm$~1.6~$\pm$~2.3~\% (ATLAS) for a theoretical prediction of 0.6\%.

\subsection{Other top-quark properties}
ATLAS measured the W helicity in \ttbar\ single-lepton and di-lepton decays~\cite{atlaswhelicity}. No significant deviations from NNLO QCD predictions were observed. A combination of the measurements in the single-lepton and di-lepton channels with the right-handed
helicity fraction set to zero leads to: $F_{0}~=~0.75~\pm~0.08~(stat.+syst.)$ and
$F_{L}~=~0.25~\pm~0.08~(stat.+syst.)$\footnote{The W bosons are produced as real
particles in top decays and their polarisation can be longitudinal, left-handed or right-handed. The fractions of events with a particular polarisation, $F_{0}$, $F_{L}$ and $F_{R}$, respectively, are referred to as helicity fractions}. ATLAS also measured the top-quark charge~\cite{atlastopcharge} and the \ttbar\ spin correlation in di-lepton decays~\cite{atlasttbarspin}.

\begin{figure}
\begin{minipage}[b]{0.49\linewidth}
\centerline{\includegraphics[width=\columnwidth]{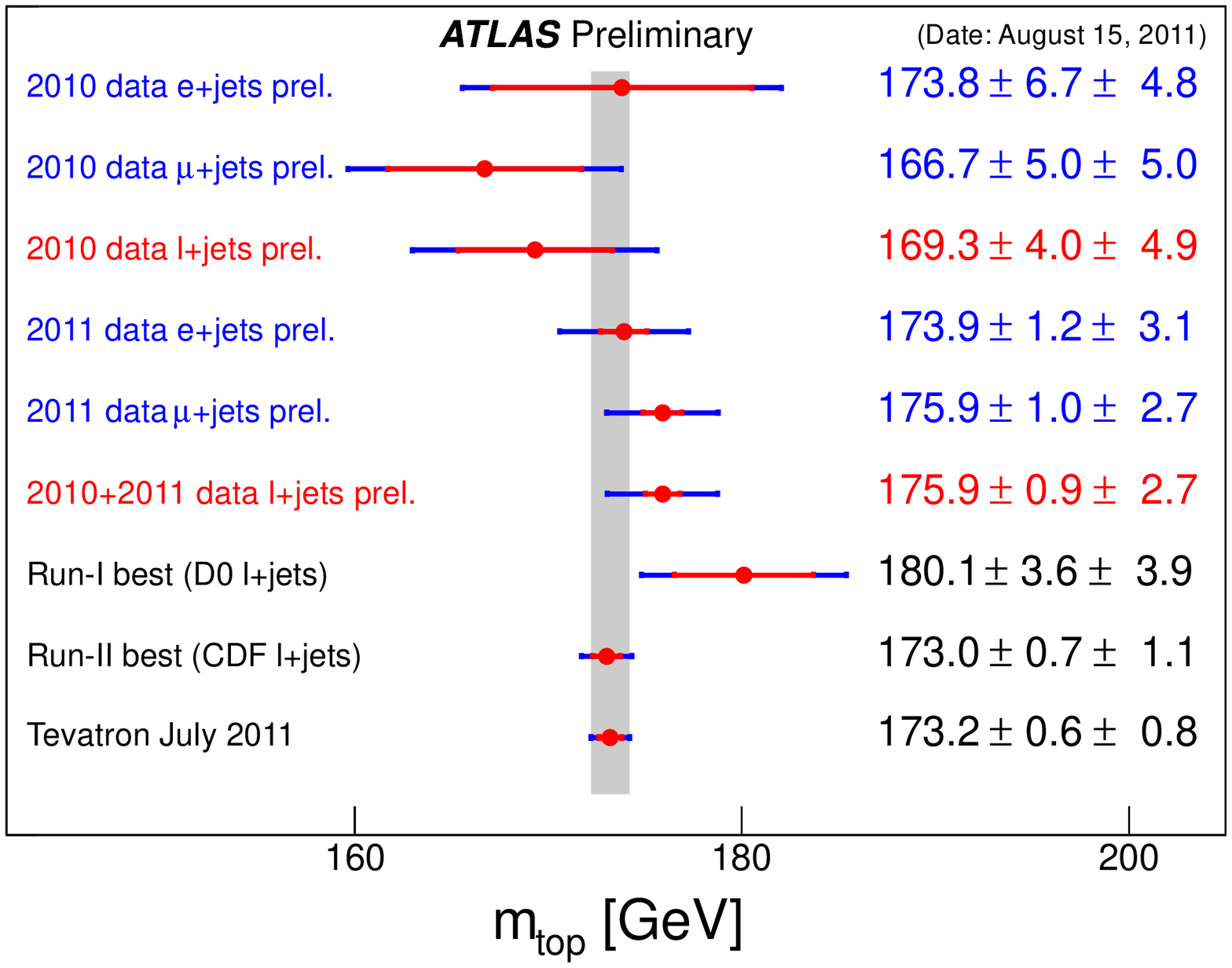}}
\caption{ Summary of the top-quark mass measurements from ATLAS Collaboration~\cite{atlasmass}.}\label{fig:mass-atlas}
\end{minipage}
\hspace{0.5cm}
\begin{minipage}[b]{0.49\linewidth}
\centerline{\includegraphics[width=\columnwidth]{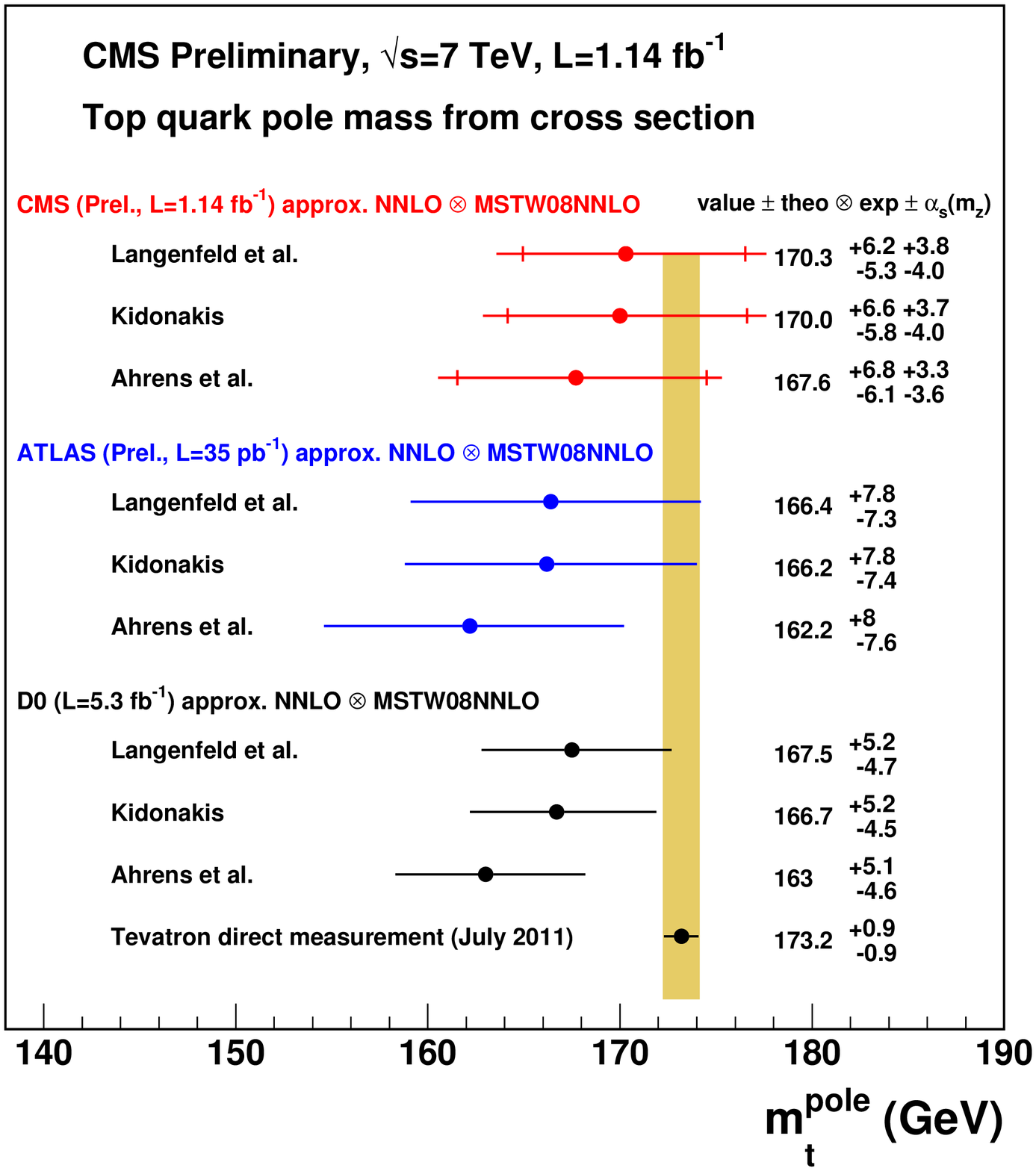}}
\caption{ Summary of the top-quark pole mass measurements from CMS, ATLAS and D0 Collaborations~\cite{cmsmpole}.}\label{fig:mass-cms}
\end{minipage}
\end{figure}

\section{New physics searches in the top-quark sector}
Several New Physics (NP) scenarios can produce deviations from the SM in \ttbar\ and single-top-quark production. Examples of NP in the top sector are: Flavour Changing Neutral Current (FCNC) decays of the top-quark to a quark with same charge and different flavour: $t~\to~(Z,\gamma,g)~q$; heavy neutral particles decaying to \ttbar, like heavy vector-bosons or Kaluza-Klein resonances: $(Z',g_{KK})~\to~\ttbar$; heavy top-like partners decaying to a top-quark and a stable or unstable neutral particles: $T~\to~t~(A^{0},Z)$.

\subsection{FCNC}
Searches of FCNC were performed by CMS with 35~\ipb and by ATLAS with 0.7~\ifb and 35~\ipb of data. CMS searched for same-sign $tt$ pairs~\cite{cmsfcnc}, that can be induced by t-channel exchange of a massive neutral vector boson ($Z'$). CMS placed a 95\% confidence level limit on the four-fermions contact interaction term for a $Z'$ mass of 2 TeV: $\frac{C_{RR}}{\Lambda}<$~2.7~TeV$^{-2}$\footnote{ $\frac{C_{RR}}{\Lambda}<$ defines the coupling strength of the four-fermions contact interaction as a function of the NP energy scale ($\Lambda$). }.
ATLAS placed 95\% confidence level limits on BR($t~\to~qZ$) $<$ 1.1\%~\cite{atlasfcnc} and $\sigma_{qg \to t} \times$~BR($t~\to~Wb$) $<$ 17.3 pb~\cite{atlasfcnc2}.

\subsection{Heavy reasonances}
Searches for heavy neutral particles decaying to \ttbar\ have been performed by ATLAS in the single-lepton final state with 0.2~\ifb of data~\cite{atlasressinglelep} and in the di-lepton final state with 1.0~\ifb of data~\cite{atlasresdilep}.  95\% confidence level limits on $g_{KK}$ mass have been placed at 0.65 TeV and 0.84 TeV respectively.
CMS searched for heavy neutral particles decaying to \ttbar\ in the $\mu$+jets and full hadronic final state with 1.1~\ifb and 0.9~\ifb of data respectively~\cite{cmsresonelep,cmsresfullhad}. 1 pb limits on the cross-section  were put with 95\% confidence level for $m_{Z'}<$~1.3 TeV in the  $\mu$+jets channel and for  $m_{Z'}<$~1.1 TeV in the full hadronic channel.

\subsection{Top partners}
Searches for top-like partners decaying as $T~\to~t~Z$ were performed by CMS in the final state with 3 leptons~\cite{cmsT} with 1.1 \ifb of data. A 95\% confidence level limit on the T mass has been set: $m_{T}>475$ GeV, as shown in Figure~\ref{fig:ttbarZ-cms}. ATLAS searched for  top-like partners decaying as $T~\to~t~A^{0}$ in \ttbar+large \MET events in the single lepton channel with 1.0 \ifb of data~\cite{atlasT}. A limit at 95\% confidence level for a production rate of 1.1 pb was set for $(m_{T},m_{A^{0}})=$ (420 GeV, 10 GeV) as shown in Figure~\ref{fig:ttbarmet-atlas}.

\begin{figure}
\begin{minipage}[b]{0.49\linewidth}
\centerline{\includegraphics[width=0.95\columnwidth]{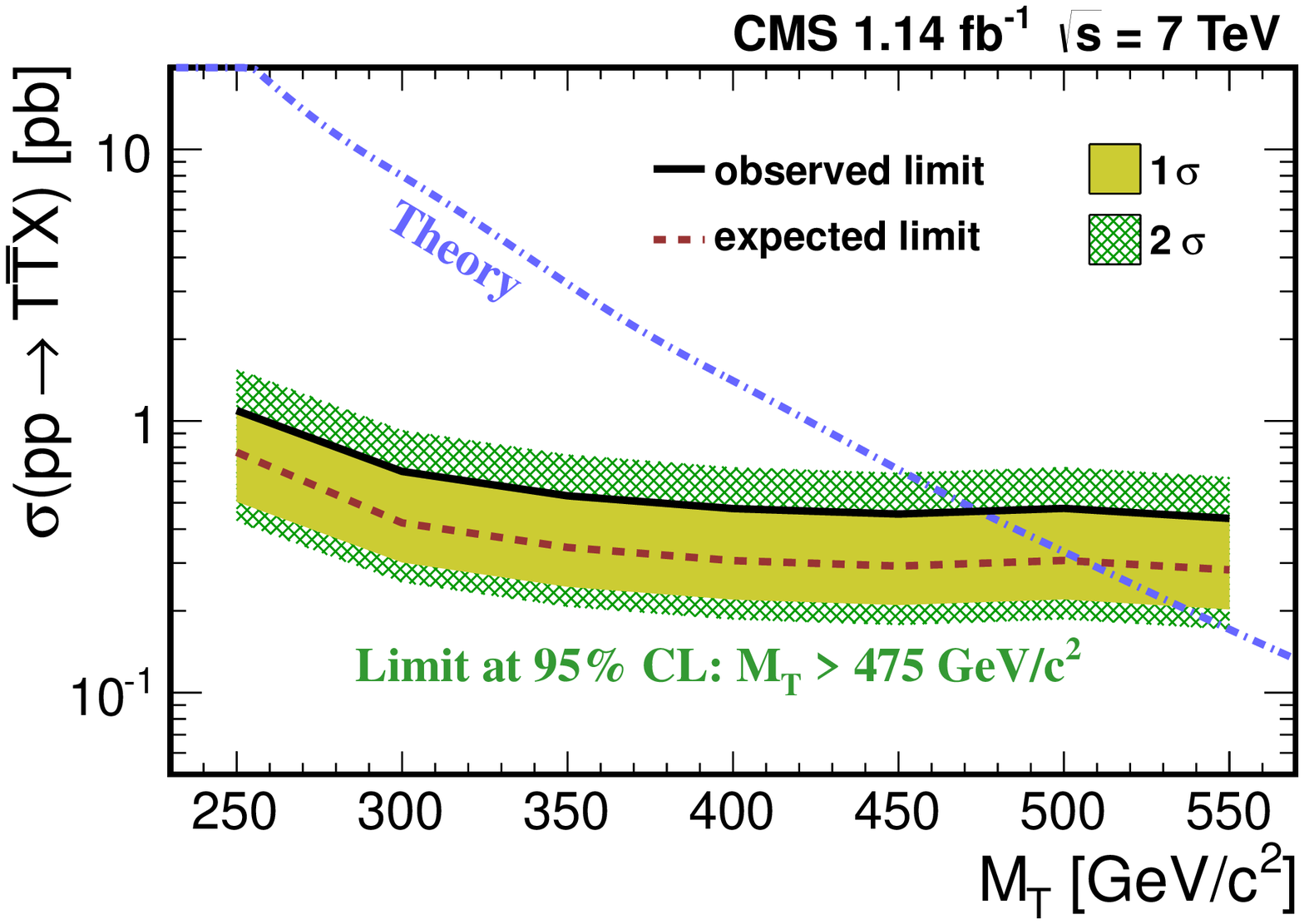}}
\caption{ Limits on the cross-section production rates of $T \to t Z$ in events with three leptons, jets and \MET\ by the CMS Collaboration~\cite{cmsT}.}\label{fig:ttbarZ-cms}
\end{minipage}
\hspace{0.5cm}
\begin{minipage}[b]{0.49\linewidth}
\centerline{\includegraphics[width=0.95\columnwidth]{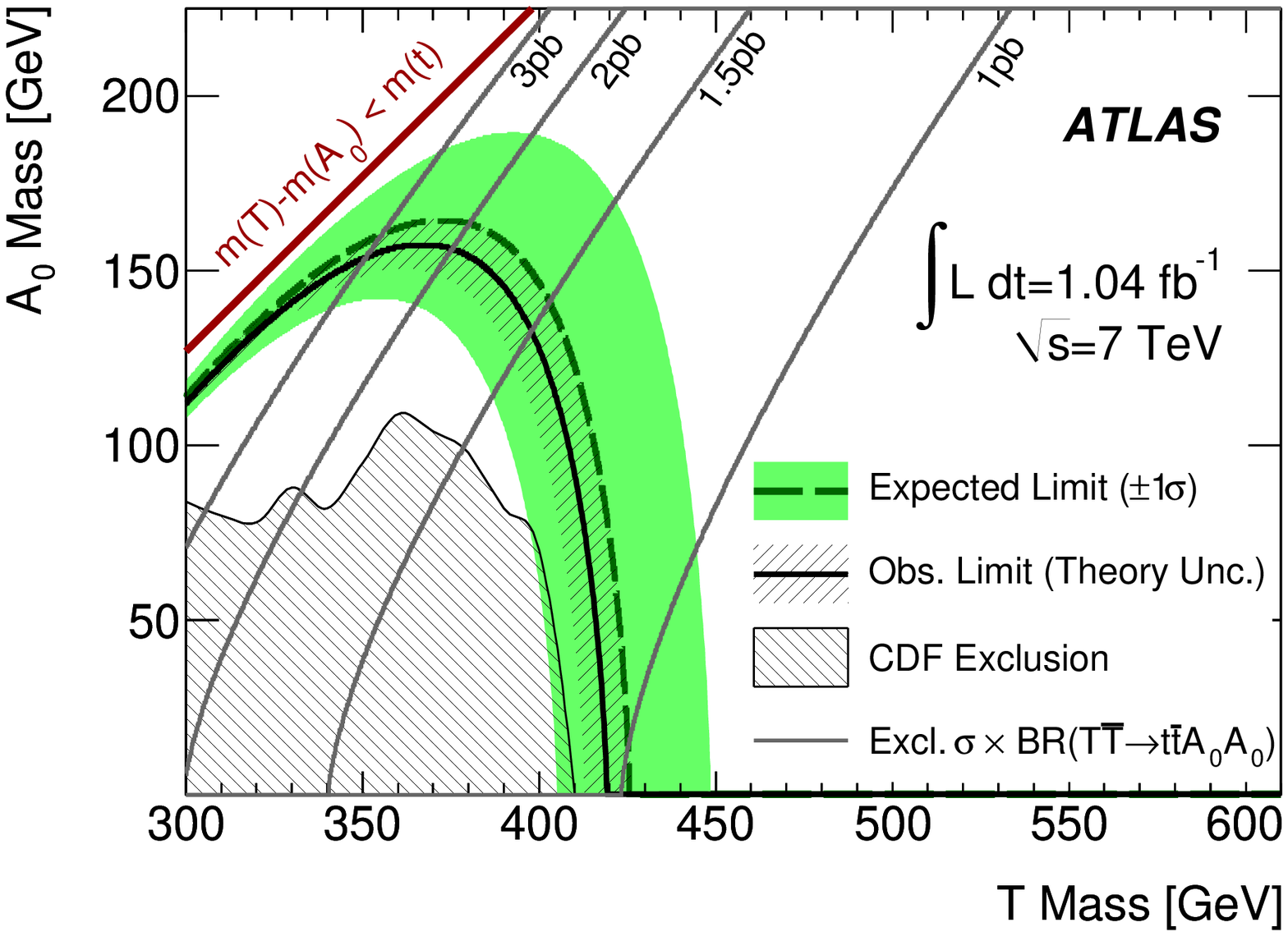}}
\caption{ Limits on the cross-section production rates of $T \to t A^{0}$ in events with one lepton, jets and large \MET\ by the ATLAS Collaboration~\cite{atlasT}.}\label{fig:ttbarmet-atlas}
\end{minipage}
\end{figure}

\section{Conclusions}
The top-quark physics program at LHC is extremely vast and complete.\\
The \xsectt\ has been measured in almost all the final states predicted by SM with an accuracy, limited by systematics, that is similar to the NNLO theoretical predictions, entering in a phase of precise measurements just one year after the first top-quark observations at the LHC. The single-top-quark production mechanism has been clearly established by LHC experiments in the $t$-channel, while it needs more data to be observed in $Wt$ channel and $s$-channel.\\
A large number of the top-quark properties has been  measured by the ATLAS and CMS experiments. The top mass, top-anti-top mass difference, $W$ helicity, \ttbar\ charge asymmetry, top charge and \ttbar\ spin correlation have been successfully measured.\\
The unprecedented energy in the centre of mass of the LHC and its rapidly increasing luminosity is delivering new stringent limits in the search for new physics related to the top-quark sector. These searches will benefit from higher collected statistics.





\begin{footnotesize}


\end{footnotesize}


\end{document}